\documentclass[traditabstract,epsf,openany]{aa}

\PassOptionsToPackage{hyphens}{url}\usepackage[colorlinks = true, breaklinks]{hyperref}
\hypersetup{urlcolor=black,linkcolor=black,citecolor=black,colorlinks=true}
\usepackage{breakurl}
\usepackage{subfigure}
\usepackage{hyperref}

\DeclareTextSymbol{\degre}{OT1}{23}

\usepackage{amsmath}
\usepackage{graphicx}
\usepackage{epstopdf}
\usepackage{txfonts}
\usepackage{natbib}
\usepackage{verbatim}
\usepackage{epstopdf}
\usepackage[flushleft]{threeparttable}
\usepackage{threeparttable}

\begin{document}

   \title{TRAPPIST photometry and imaging monitoring of comet C/2013 R1(Lovejoy): Implications for the origin of daughter species}

\author{
          C.~Opitom\inst{1},
          E.~Jehin\inst{1},
          J.~Manfroid\inst{1},
          D.~Hutsem\'ekers\inst{1},
          M.~Gillon\inst{1},
          P.~Magain\inst{1}
       }

\offprints{cyrielle.opitom@ulg.ac.be}
\institute{ 
            $^1$ Institut d'Astrophysique et G\'eophysique, Universit\'{e} de Li\`{e}ge, all\'{e}e du 6 Ao\^{u}t 17, B-4000 Li\`{e}ge, Belgium\\  
          }
\date{Received date / accepted date}
\authorrunning{C. Opitom et al.}
\titlerunning{Comet C/2013 R1 (Lovejoy)}

  \abstract
  {
We report the results of the narrow-band photometry and imaging monitoring of comet C/2013 R1 (Lovejoy) with the robotic telescope TRAPPIST (La Silla observatory). We gathered around 400 images over 8 months pre- and post-perihelion between September 12, 2013 and July 6, 2014. We followed the evolution of the OH, NH, CN, $\mathrm{C_{3}}$, and $\mathrm{C_{2}}$ production rates computed with the Haser model, as well as the evolution of the dust production. All five gas species display an asymmetry about perihelion, since the rate of brightening is steeper than the rate of fading. The study of the coma morphology reveals gas and dust jets that indicate one or several active zone(s) on the nucleus. The dust, $\mathrm{C_{2}}$, and $\mathrm{C_{3}}$ morphologies present some similarities, while the CN morphology is different. OH and NH are enhanced in the tail direction. The study of the evolution of the comet activity shows that the OH, NH, and $\mathrm{C_{2}}$ production rate evolution with the heliocentric distance is correlated to the dust evolution. The CN and, to a lesser extent, the $\mathrm{C_{3}}$ do not display such a correlation with the dust. This evidence and the comparison with parent species production rates indicate that $\mathrm{C_{2}}$ and $\mathrm{C_{3}}$, on one hand, and OH and NH, on the other, could be -at least partially- released from organic-rich grains and icy grains. On the contrary, all evidences point to HCN being the main parent of CN in this comet.
  }

   \keywords{Comets: general, Comets: individual: C/2013 R1 (Lovejoy)
               }

   \maketitle


\section{Introduction}
\label{intro}

\indent
Comets are remnants of the early solar system. They have undergone little alteration since their formation, and their nuclei are thought to preserve physical and chemical information about the early solar nebula. When a comet approaches the Sun, the ices contained in the nucleus sublimate, and complex molecules are released into the coma. These molecules, the parent species, can be observed in the infrared and radio domains. In optical studies we observe daughter or even granddaughter species (CN, $\mathrm{C_{3}}$, $\mathrm{C_{2}}$, etc.) that are the product of the parent species' photodissociation. Since the relationship between parent, daughter, and granddaughter volatiles is still not understood well, it remains very difficult to link the radicals observed in the optical to nucleus ice abundances. Recent studies comparing the abundances of parent volatiles derived from infrared observations and abundances of daughter species derived from optical spectroscopy or photometry reveal a certain diversity among comets (\citealt{Mumma2011}).
 
\indent
The case of CN is a good illustration of the problems encountered while trying to link parent and daughter species. CN was first thought to be the product of HCN photodissociation. However, studies have shown that for some comets, the HCN abundance was insufficient to account for the CN production rates (\citealt{Fray2005}, \citealt{DelloRusso2009}). The same discrepancies were also noted between $\mathrm{C_{2}H_{2}}$ and $\mathrm{C_{2}}$ production rates. Moreover, the spatial distribution of $\mathrm{C_{2}}$ and CN in the coma is sometimes not consistent with these two species being the product of $\mathrm{C_{2}H_{2}}$ and HCN photodissociation. It has then been suggested that these radicals could have other parents or could be released by organic-rich grains in the coma (\citealt{AHearn1986}, \citealt{Combi1997}). However, additional parents for these two species are not needed for all comets. Since the sample of comets observed simultaneously in the infrared and the optical is still scarce, it remains difficult to have a comprehensive picture of the relationship between parent and daughter species.

\indent
We present here the monitoring of comet C/2013 R1 (Lovejoy) performed with TRAPPIST (\citealt{Jehin2011}). We followed the evolution of the production rates of five daughter species and their coma morphology. We explore the link between the comet activity evolution and morphology and the possible origin of the observed radicals. Comet C/2013 R1 (Lovejoy) is a dynamically old long-period comet discovered on September 7, 2013 at 1.97 au from the Sun (\citealt{Guido2013}). It was first detected by the Australian amateur astronomer Terry Lovejoy at magnitude 14.4 and brightened up to magnitude 4.5. The comet reached perihelion on December 22, 2013 at 0.8 au from the Sun. It follows an inclined and highly eccentric orbit ($i=64\degre$ and $e=0.9924$). After the disintegration of comet ISON in late November, comet Lovejoy became an ideal backup target, and it has been extensively observed with a variety of instruments at different wavelengths around its perihelion passage. Thanks to the fast response possible with TRAPPIST, we were able to already observe the comet a few days after its discovery. Since the comet was also observable post-perihelion, it was an extremely interesting target for the study of its activity and chemical composition, and we started an intensive monitoring with narrow-band filters.

\section{Observations and data reduction}
\label{observations}
\indent
We observed comet Lovejoy with the TRAPPIST 60-cm robotic telescope at La Silla observatory (\citealt{Jehin2011}). TRAPPIST is equipped with a $2K\times2K$ thermoelectrically cooled FLI Proline CCD camera with a field of view of 22\textquoteright $\times$ 22\textquoteright. We binned the pixels 2 by 2 and obtained a resulting plate scale of $1.302$\textquotedblright/pixel. We observed the comet with HB narrow band filters \citep{Filters} isolating the emission bands of $\mathrm{OH}$, $\mathrm{NH}$, $\mathrm{CN}$, $\mathrm{C_{3}}$, and $\mathrm{C_{2}}$, as well as emission free continuum at four wavelengths (see \citealt{Filters} for filters characteristics). Images have also been taken with broad band B, V, Rc, and Ic Johnson-Cousin filters.

\indent
We collected about 400 images over eight months. We started the observations on September 12, 2013 when the comet was at 1.91 au from the Sun. We observed once or twice a week until November 16, when it was too close to the Sun. Just after the perihelion, the comet was too far north to be observed from La Silla, and we recovered it on February 13, 2014. We kept observing with narrow band filters as long as the comet was bright enough, until July 6, 2014 ($r=3.10$ au, $r$ being the heliocentric distance). After April 25 ($r=2.20$ au) the signal was too low in the OH and NH filters to derive reliable production rates. Exposure times ranged from 15 to 120 seconds for broad band filters and from 60 to 900 seconds for narrow band filters. Most nights were photometric, and we discarded cloudy nights.

\begin{table}[h!]
\begin{center}
\caption[Caption for LOF]{Parent ($L_{P}$) and daughter ($L_{D}$) scalelengths used in the Haser model.}
\begin{tabular}{lll}
\hline
\hline
  Species  & $L_{P}$  & $L_{D}$\\
           & (km)     & (km))  \\
\hline
\hline
  OH & $2.40$ $ 10^{4}$ & $1.60$ $ 10^{5}$ \\
  NH & $5.00$ $ 10^{4}$ & $1.50$ $ 10^{5}$ \\
  CN & $1.30$ $ 10^{4}$ & $2.10$ $ 10^{5}$ \\
  $\mathrm{C_{3}}$ & $2.80$ $ 10^{3}$ & $2.70$ $ 10^{4}$ \\
  $\mathrm{C_{2}}$ & $2.20$ $ 10^{4}$ & $6.60$ $ 10^{4}$ \\
\hline
\hline
\end{tabular}
\label{scalelengths}
\tablefoot{
These values are given for 1 au and scale as $\mathrm{r^{2}}$ (\citealt{AHearn1995}).}
\end{center}
\end{table}

\indent
Calibration followed standard procedures using frequently updated master bias and flat and dark frames. The removal of the sky contribution and the photometric calibration of the images followed the procedure described in \citet{Opitom2015}. After calibration, we derived a median radial profile for each image. We then removed the dust contamination in the gas radial profiles by subtracting a properly scaled dust profile. From the $\mathrm{OH}$, $\mathrm{NH}$, $\mathrm{CN}$, $\mathrm{C_{3}}$, and $\mathrm{C_{2}}$ radial profiles, we derived production rates by adjusting a radial coma brightness model. The model adjustment is performed at a nucleocentric distance around 10~000 km to avoid PSF and seeing effects around the optocenter. At greater nucleocentric distances, the signal usually becomes fainter, especially in the OH filter for which the signal-to-noise ratio is lower. Fluorescence efficiencies (also called g~factors) from David Schleicher's website\footnote{Link: \url{http://asteroid.lowell.edu/comet/gfactor.html}} were used to convert fluxes into column densities. $\mathrm{C_{2}}$ g~factors only consider the $\mathrm{C_{2}}$ in triplet state \citep{AHearn1982}. $\mathrm{C_{3}}$ g-factors are also from \cite{AHearn1982}. CN and NH fluorescence efficiencies vary with both the heliocentric distance and velocity and are respectively taken from \cite{Schleicher2010} and from \cite{Meier1998}. For OH g~factors, which also vary with the heliocentric velocity, we considered unquenched values of the ground state \citep{Schleicher1988}. All fluorescence efficiencies are scaled as $r^{2}$. We used the Haser model to compute gas production rates \citep{Haser1957}. This model is fairly simple and considers the photodissociation of mother molecules into daughter and granddaughter molecules in a spherically symmetric coma. The model has only two free parameters: the mother and daughter scalelengths. We used the scalelengths from \cite{AHearn1995}. These are shown in Table \ref{scalelengths} for a comet at 1 au from the Sun and are scaled as $r^{2}$. We used a constant velocity of 1 km$/$s for both parent and daughter species, as in \citet{AHearn1995}.

\begin{figure}[h!]
\centering
\includegraphics[width=7.5cm]{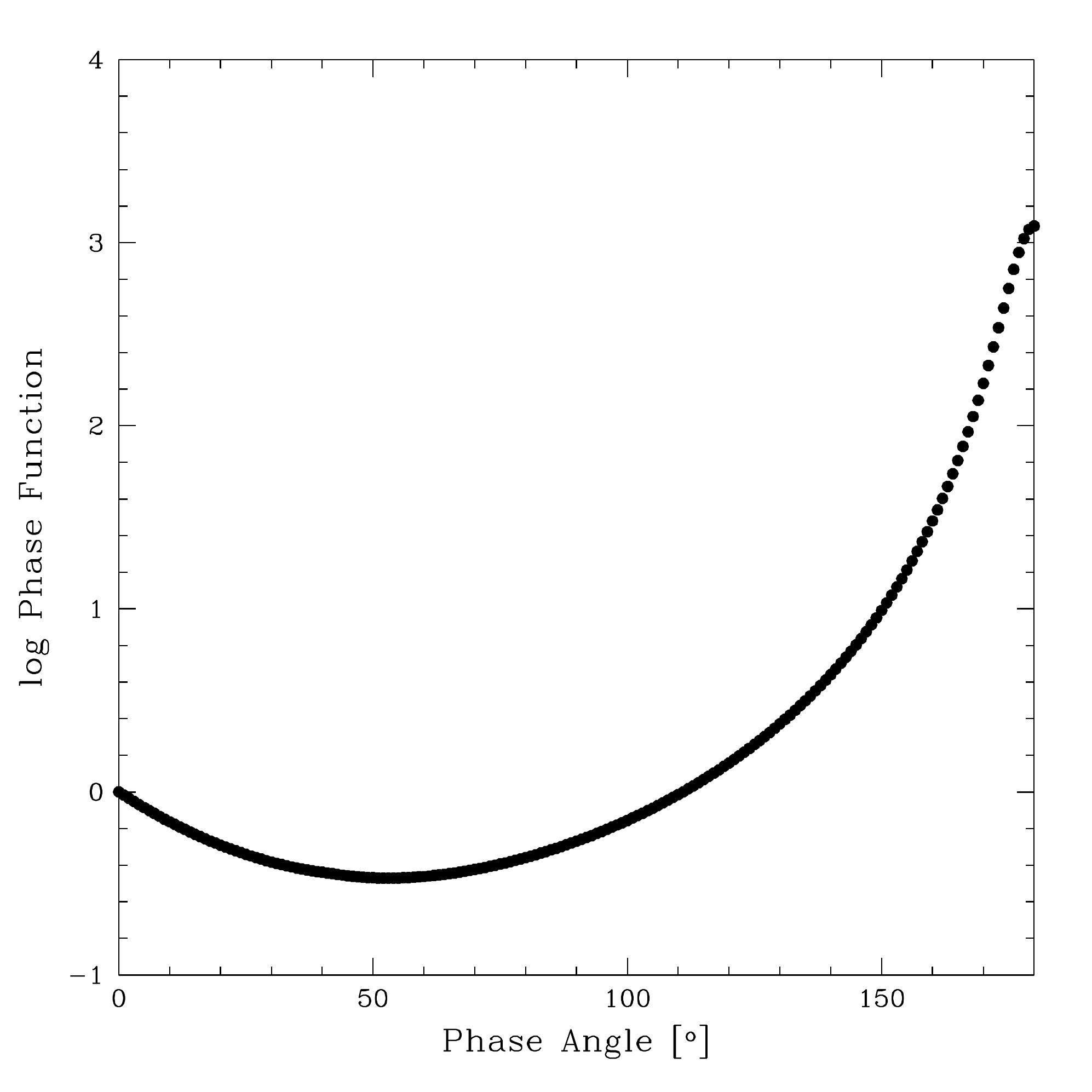}
\caption{Phase function used to correct the $A(\theta) f\rho$ from the phase angle effect.}
  \label{phasefunc}
\end{figure}

\indent
From the profiles in the dust filters, we derived the $A(\theta) f\rho$ parameter at three wavelengths. This quantity, first introduced by \cite{AHearn1984}, is representative of the dust activity and independent of the aperture size if the dust coma is in a steady state (following a $1/\rho$ radial profile, $\rho$ being the nucleocentric distance). Since the phase angle was varying significantly during the observations (from almost 70\degre$ $ around mid-November 2013 to only a few degrees in early July 2014), we corrected the $A(\theta) f\rho$ from the phase angle effect to obtain $A(0) f\rho$. Several phase functions have been successively proposed to correct the $A(\theta) f\rho$ for the phase angle effect (\citealt{Divine1981}, \citealt{Hanner1989}, \citealt{Schleicher1998} or \citealt{Marcus2007}). We used the phase function described by David Schleicher\footnote{Link: \url{http://asteroid.lowell.edu/comet/dustphase.html}}, which is represented in Fig. \ref{phasefunc}. It is a composite of two different phase functions from \citet{Schleicher1998} and \citet{Marcus2007}. We normalized the $A(\theta) f\rho$ values at a 0\degre$ $ angle.

\section{Data analysis and results}
\label{analysis}

\indent
In this section, we present the evolution of comet Lovejoy activity and composition with heliocentric distance. We compare the activity before and after the perihelion passage and also discuss the morphology of the dust and gas coma.

\subsection{Production rates evolution}
\label{rate}

\begin{table*}[]
\centering
\caption{$\mathrm{OH}$, $\mathrm{NH}$, $\mathrm{CN}$, $\mathrm{C_{3}}$, and $\mathrm{C_{2}}$ production rates and $Af\rho$ for comet C/2013 R1 (Lovejoy)}
\begin{tabular}{p{2.05cm}p{0.35cm}p{0.40cm}p{1.93cm}p{1.93cm}p{1.93cm}p{1.93cm}p{1.93cm}p{1.83cm}}
  \hline
  \hline
  UT date & $\mathrm{r}$ & $\Delta$ &  \multicolumn{5}{c}{Production rates ($10^{25}$ mol/s)} & $A(0) f\rho$   \\
   & (au) & (au) & Q(OH)  & Q(NH) & Q(CN) & Q($\mathrm{C_{3}}$) & Q($\mathrm{C_{2}})$ & ($10^{2}$ cm) \\
  \hline
  \hline
2013 Sep 12.38 & 1.91 & 1.81 & 870$\pm$135  &                &                & 0.67$\pm$0.04 & 2.35$\pm$0.13  &               \\        
2013 Sep 16.39 & 1.85 & 1.71 &              &                & 3.14$\pm$0.12  &               & 2.84$\pm$0.20  & 3.79$\pm$0.28 \\    
2013 Sep 21.33 & 1.79 & 1.59 & 823$\pm$144  & 4.89$\pm$1.63  & 3.17$\pm$0.14  & 0.74$\pm$0.06 & 3.10$\pm$0.21  & 4.37$\pm$0.37 \\
2013 Sep 24.39 & 1.75 & 1.52 &              &                & 3.67$\pm$0.12  &               &                &               \\
2013 Sep 25.35 & 1.73 & 1.50 &              &                & 3.70$\pm$0.14  &               & 4.32$\pm$0.15  & 4.04$\pm$0.39 \\
2013 Oct 04.35 & 1.61 & 1.28 &              & 9.73$\pm$1.13  & 3.93$\pm$0.12  &               &                &               \\
2013 Oct 05.33 & 1.60 & 1.25 & 1120$\pm$180 &                & 4.28$\pm$0.13  & 1.03$\pm$0.04 & 5.29$\pm$0.15  & 6.30$\pm$0.18 \\
2013 Oct 08.33 & 1.56 & 1.18 & 1520$\pm$180 &                & 3.74$\pm$0.12  &               & 4.49$\pm$0.16  & 6.85$\pm$0.24 \\
2013 Oct 10.31 & 1.53 & 1.13 & 1470$\pm$210 &                & 5.02$\pm$0.16  &               & 5.26$\pm$0.42  & 7.52$\pm$0.24 \\
2013 Oct 20.29 & 1.40 & 0.89 & 2160$\pm$330 &                & 5.67$\pm$0.23  &               & 7.45$\pm$0.34  & 10.00$\pm$0.50 \\
2013 Oct 21.33 & 1.38 & 0.87 & 2380$\pm$250 & 15.80$\pm$1.70 & 6.38$\pm$0.19  & 1.19$\pm$0.04 & 7.63$\pm$0.21  &                \\
2013 Oct 24.32 & 1.34 & 0.80 & 2720$\pm$290 & 19.10$\pm$1.80 & 6.96$\pm$0.21  & 1.85$\pm$0.08 & 9.10$\pm$0.34  & 13.10$\pm$0.30 \\
2013 Oct 26.30 & 1.32 & 0.75 & 2860$\pm$330 & 20.00$\pm$2.00 & 7.20$\pm$0.31  & 1.93$\pm$0.07 & 9.62$\pm$0.39  & 14.20$\pm$0.40 \\
2013 Nov 02.35 & 1.23 & 0.61 &              &                & 7.97$\pm$0.27  & 1.92$\pm$0.07 & 12.20$\pm$0.30 &                \\
2013 Nov 03.37 & 1.21 & 0.59 & 3010$\pm$350 & 23.70$\pm$1.70 &                &               &                &                \\
2013 Nov 13.34 & 1.09 & 0.43 & 3240$\pm$790 & 31.10$\pm$4.60 & 10.10$\pm$0.70 &               & 16.10$\pm$0.50 & 22.60$\pm$0.80 \\
2013 Nov 16.34 & 1.06 & 0.41 &              & 43.80$\pm$8.60 & 12.90$\pm$1.20 & 3.40$\pm$0.31 & 20.60$\pm$0.09 & 28.50$\pm$2.10 \\
2014 Feb 13.39 & 1.26 & 1.56 &                          &                & 8.76$\pm$0.44  &               & 13.70$\pm$0.50 & 42.00$\pm$1.70 \\
2014 Feb 15.40 & 1.28 & 1.56 & 1630$\pm$320 & 20.80$\pm$3.50 &                &               & 11.30$\pm$0.40 & 34.90$\pm$1.40 \\
2014 Feb 16.40 & 1.30 & 1.57 &              &                & 6.12$\pm$0.36  &               &                &                \\
2014 Feb 21.39 & 1.36 & 1.58 & 1380$\pm$230 & 17.50$\pm$2.60 & 5.97$\pm$0.26  & 2.10$\pm$0.11 & 8.63$\pm$0.31  & 28.30$\pm$0.90 \\
2014 Feb 22.39 & 1.37 & 1.58 &              &                &                &               & 9.59$\pm$0.27  & 29.10$\pm$1.00 \\
2014 Feb 28.39 & 1.45 & 1.59 &              &                &                &               &                & 21.70$\pm$0.90 \\
2014 Mar 01.39 & 1.47 & 1.59 &              &                &                &               &                & 24.00$\pm$0.90 \\
2014 Mar 06.39 & 1.54 & 1.59 & 1320$\pm$160 & 14.30$\pm$1.70 & 5.10$\pm$0.17  & 1.69$\pm$0.06 & 7.10$\pm$0.17  & 21.80$\pm$0.60 \\
2014 Mar 13.39 & 1.63 & 1.58 & 1160$\pm$130 & 11.90$\pm$1.50 & 4.45$\pm$0.16  & 1.46$\pm$0.06 & 5.87$\pm$0.17  & 17.80$\pm$0.60 \\
2014 Mar 17.39 & 1.68 & 1.57 &              &                &                &               &                & 15.20$\pm$1.20  \\
2014 Mar 18.38 & 1.70 & 1.57 & 939$\pm$115  & 9.10$\pm$1.76  & 3.99$\pm$0.17  & 1.33$\pm$0.08 & 5.97$\pm$0.26  & 12.50$\pm$0.70 \\
2014 Mar 20.34 & 1.72 & 1.56 &              &                & 4.01$\pm$0.14  &               &                &                \\
2014 Mar 25.39 & 1.79 & 1.55 &              &                & 3.80$\pm$0.12  &               & 4.66$\pm$0.21  & 13.20$\pm$0.50 \\
2014 Mar 26.38 & 1.80 & 1.55 & 859$\pm$91   & 8.88$\pm$1.16  & 3.95$\pm$0.14  & 1.25$\pm$0.05 & 4.43$\pm$0.29  & 13.10$\pm$0.40 \\
2014 Apr 10.38 & 2.00 & 1.50 & 627$\pm$68   &                & 3.40$\pm$0.14  & 0.98$\pm$0.04 & 3.26$\pm$0.19  & 9.35$\pm$0.35  \\
2014 Apr 14.42 & 2.06 & 1.49 &              &                & 2.84$\pm$0.15  &               & 2.72$\pm$0.23  & 8.92$\pm$0.48  \\
2014 Apr 15.25 & 2.07 & 1.48 & 561$\pm$56   &                &                & 1.02$\pm$0.03 &                &                \\
2014 Apr 16.42 & 2.08 & 1.48 &              & 2.93$\pm$2.45  &                &               &                &                \\
2014 Apr 18.38 & 2.11 & 1.48 & 512$\pm$60   &                & 2.72$\pm$0.15  &               & 2.91$\pm$0.24  & 7.56$\pm$0.45 \\
2014 Apr 19.34 & 2.12 & 1.47 &              & 3.79$\pm$1.69  &                & 0.72$\pm$0.07 &                &               \\
2014 Apr 22.30 & 2.16 & 1.47 & 476$\pm$56   &                & 2.93$\pm$0.13  &               & 3.07$\pm$0.30  & 7.21$\pm$0.40 \\
2014 Apr 25.38 & 2.20 & 1.47 & 413$\pm$65   &                & 2.50$\pm$0.13  &               & 2.03$\pm$0.29  & 6.70$\pm$0.33 \\
2014 May 14.16 & 2.44 & 1.50 &              &                & 2.22$\pm$0.23  &               & 1.72$\pm$0.40  & 4.54$\pm$0.76 \\
2014 May 24.34 & 2.57 & 1.57 &              &                & 1.80$\pm$0.21  &               &                & 3.96$\pm$0.39 \\
2014 May 27.32 & 2.61 & 1.60 &              &                & 1.72$\pm$0.18  & 0.68$\pm$0.05 & 1.97$\pm$0.45  & 3.66$\pm$0.39  \\
2014 May 31.18 & 2.66 & 1.64 &              &                & 1.91$\pm$0.14  & 0.71$\pm$0.07 & 1.98$\pm$0.30  &                \\  
2014 Jun 07.21 & 2.74 & 1.73 &              &                &                            &                        &                        & 3.61$\pm$0.26 \\
2014 Jun 30.21 & 3.02 & 2.15 &              &                &                            &               &                & 3.01$\pm$0.41 \\
2014 Jul 04.03 & 3.07 & 2.23 &              &                &                            &               &                         & 3.05$\pm$0.35 \\
2014 Jul 06.99 & 3.10 & 2.30 &              &                &                            &               &                         & 2.62$\pm$0.59 \\ 
    
\hline
\hline
\end{tabular}
\label{obstab}
\tablefoot{
$r$ and $\Delta$ are respectively the heliocentric and geocentric distances. The date given in the first column is the mid-time of the observations.}
\end{table*}

\begin{figure}[htbp!]
\centering
\includegraphics[width=9.0cm]{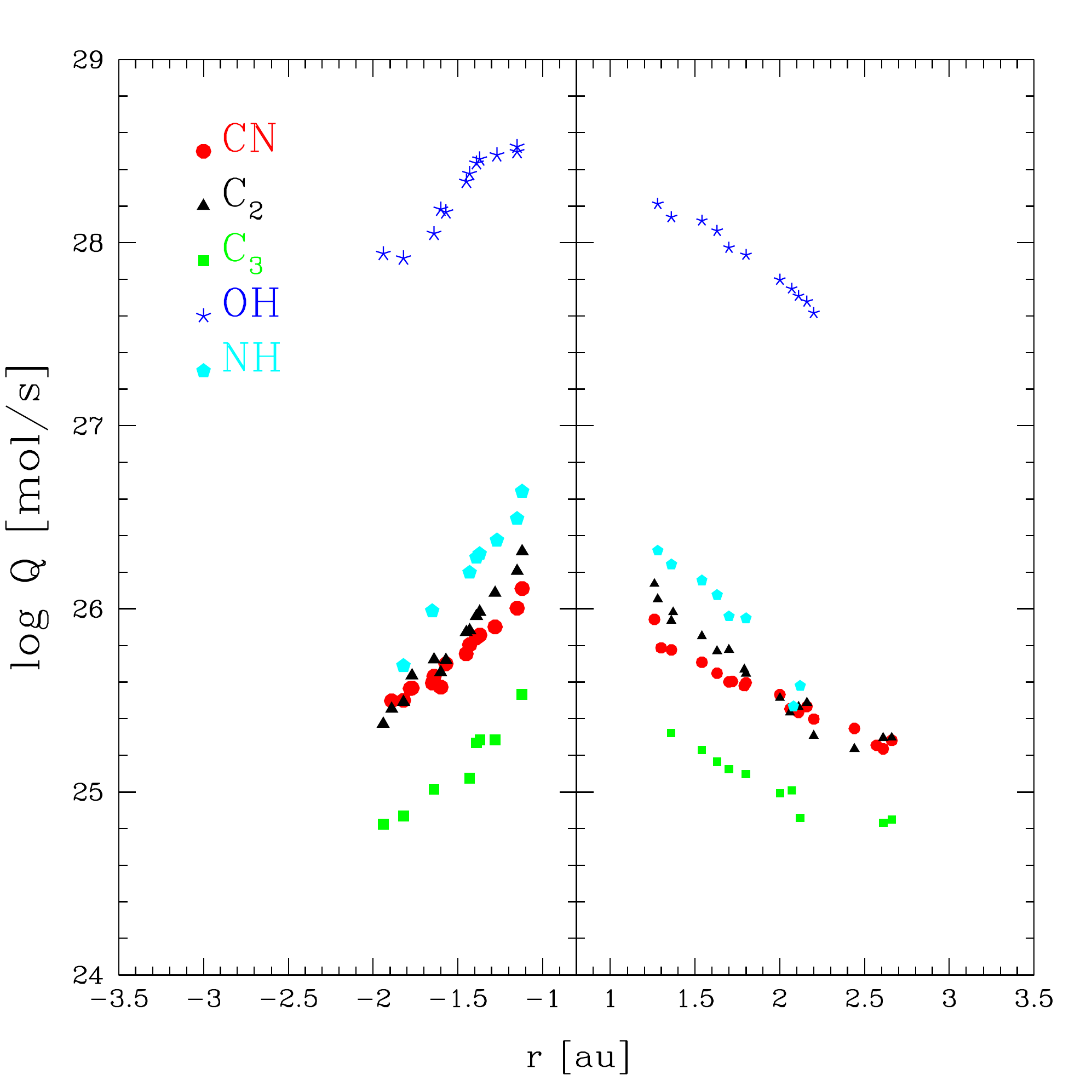}
\caption{Logarithm of the $\mathrm{OH}$, $\mathrm{NH}$, $\mathrm{CN}$, $\mathrm{C_{3}}$, and $\mathrm{C_{2}}$ production rates as a function of the heliocentric distance \textit{r} (au). The \textit{r}-axis is interrupted at perihelion ($0.8$ au).}
  \label{Qallsapin}
\end{figure}

\begin{figure*}[]
\centering
\subfigure{ \includegraphics[width=7.3cm]{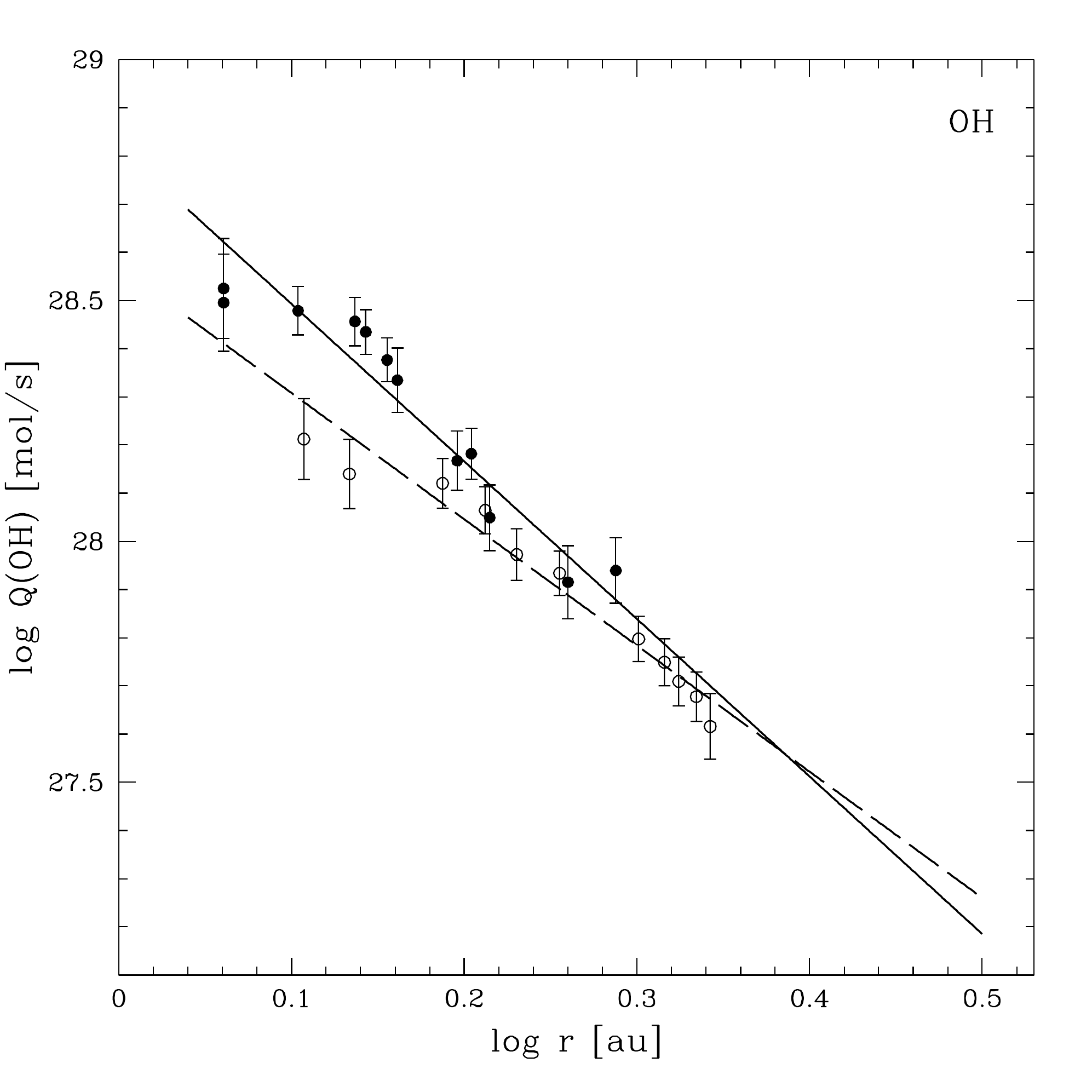}}
\hspace{2cm}
\subfigure{ \includegraphics[width=7.3cm]{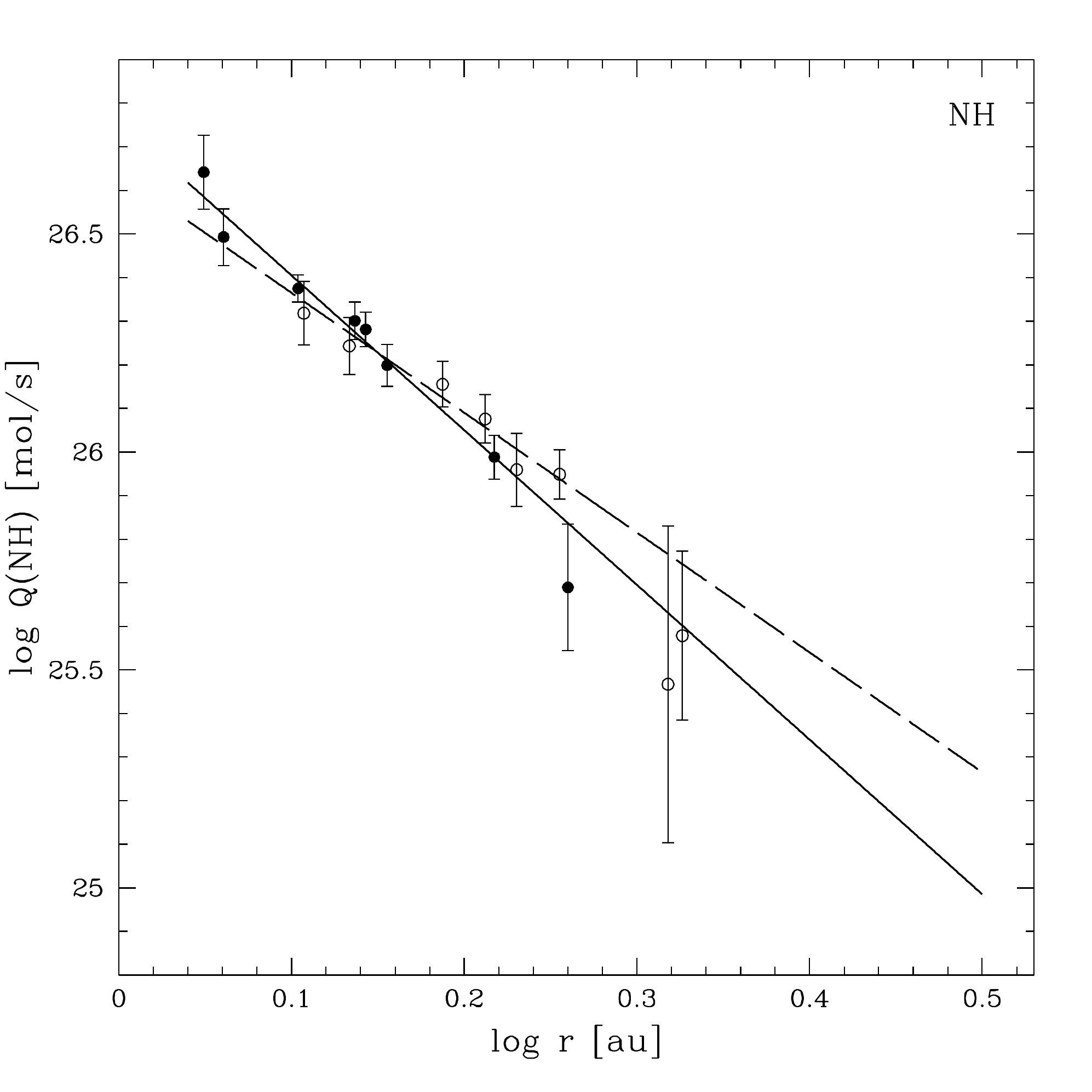}}

\subfigure{ \includegraphics[width=7.3cm]{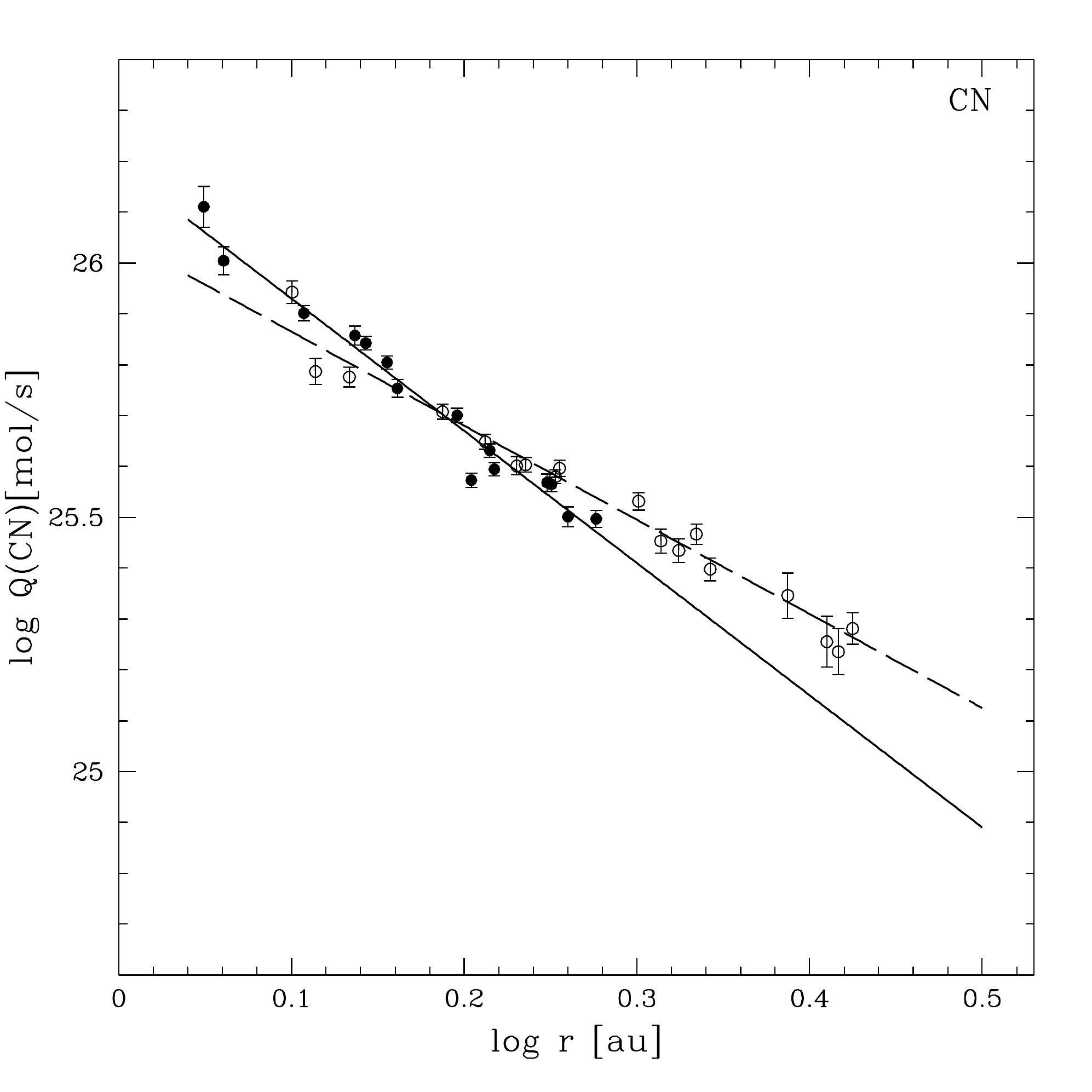}}
\hspace{2cm}
\subfigure{ \includegraphics[width=7.3cm]{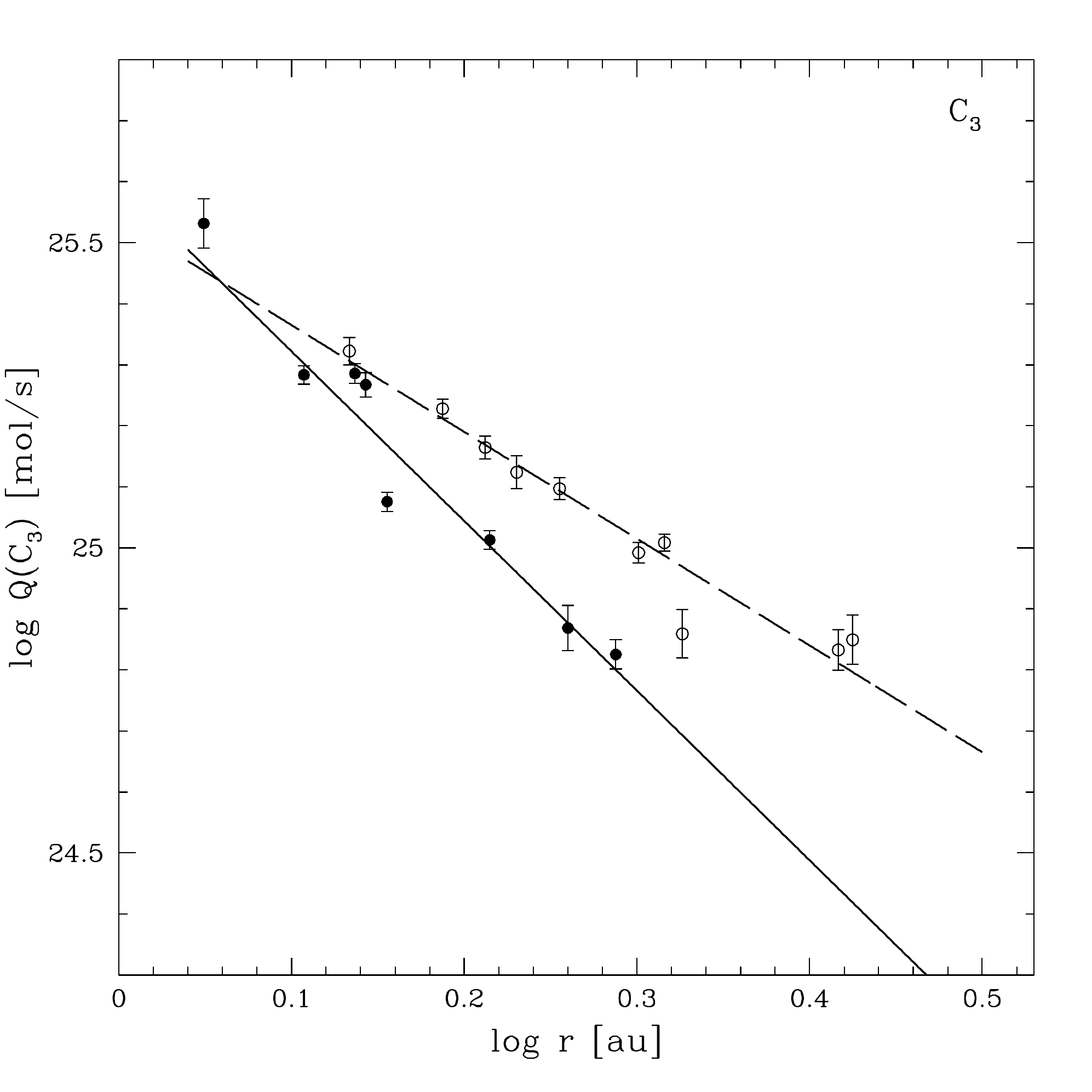}}

\subfigure{ \includegraphics[width=7.3cm]{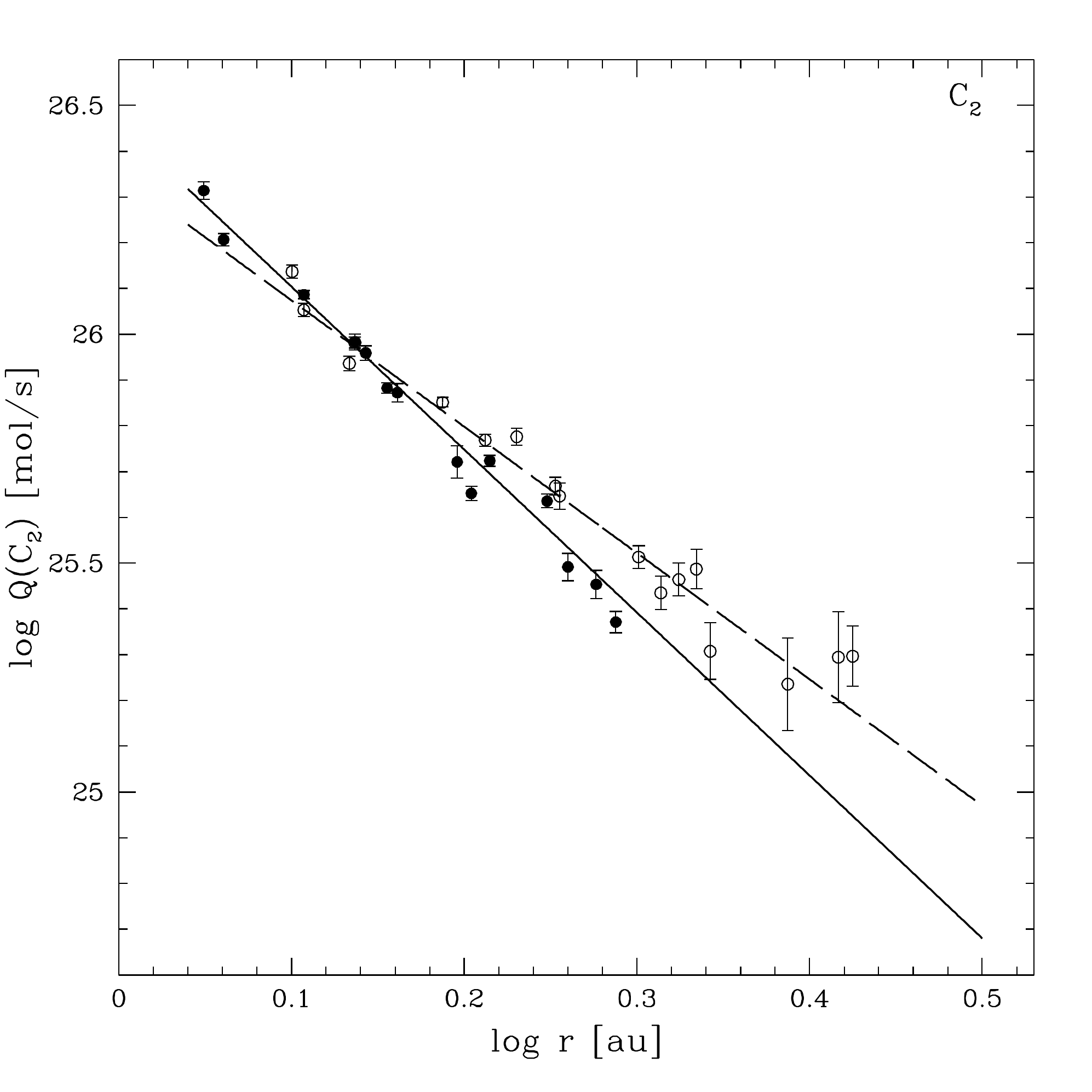}}
\hspace{2cm}
\subfigure{ \includegraphics[width=7.3cm]{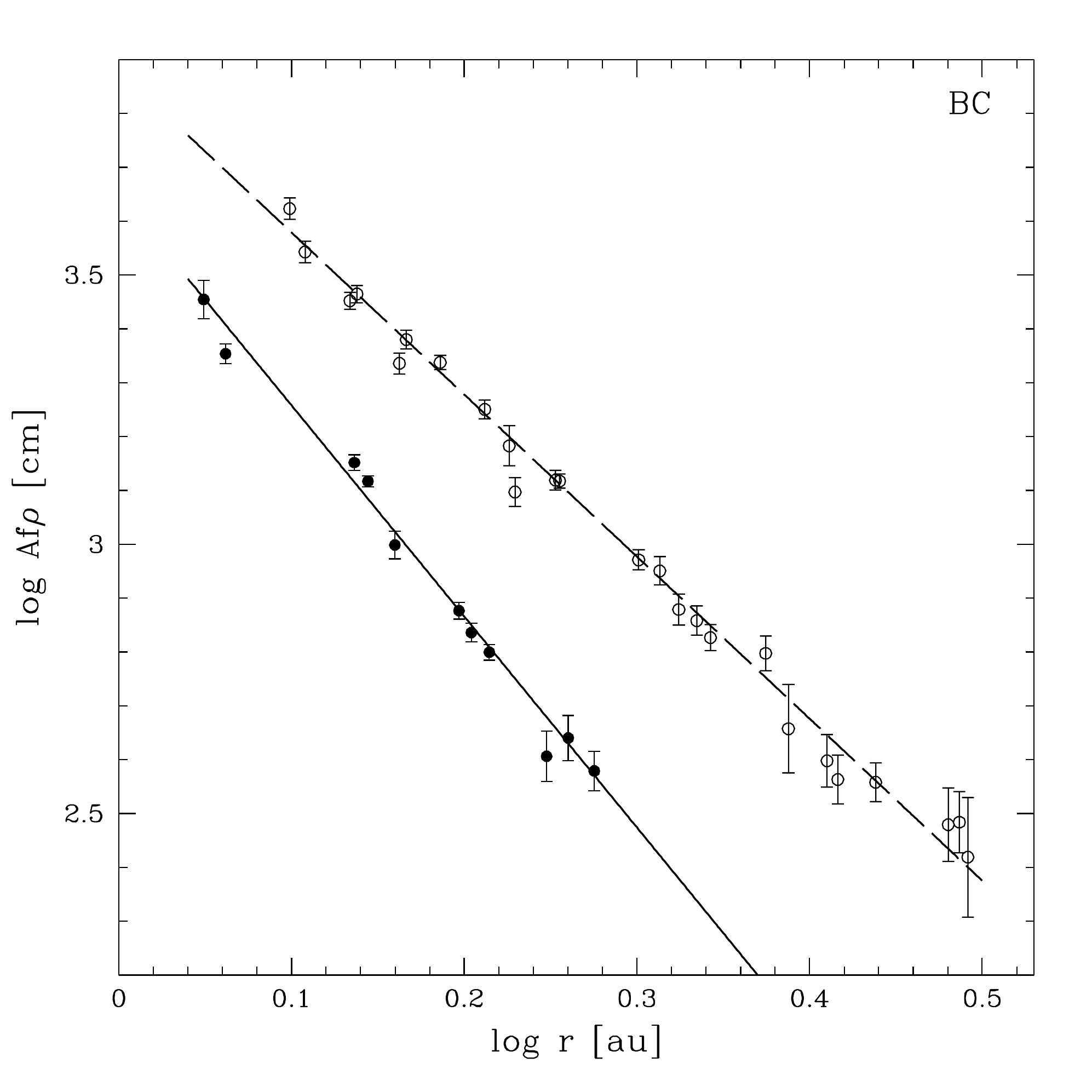}}
\caption{$\mathrm{OH}$, $\mathrm{NH}$, $\mathrm{CN}$, $\mathrm{C_{3}}$, and $\mathrm{C_{2}}$ production rates and $A(0) f\rho$ as a function of the heliocentric distance (\textit{r}). Pre-perihelion data points are represented with filled symbols, and post-perihelion data points are represented with open symbols. Full lines represent linear fits of the production rates and $A(0) f\rho$ variation with the heliocentric distance pre-perihelion. Dashed lines are the same for post-perihelion values.}
  \label{gasrh}
\end{figure*}

\indent
This section focuses on the analysis of the comet activity and composition evolution with the heliocentric distance. For each image in the narrow band filters, we computed the $A(\theta) f\rho$ value or the production rate, depending on the filter. The results are given in Table \ref{obstab}, where we give the normalized $A(0) f\rho$ value and the production rates for each image. Gas species are also represented all together  in Fig. \ref{Qallsapin}. The errors take both the errors in the flux calibration process and the sky contribution removal into account. Figure \ref{gasrh} shows the evolution of the $\mathrm{OH}$, $\mathrm{NH}$, $\mathrm{CN}$, $\mathrm{C_{3}}$, and $\mathrm{C_{2}}$ production rates and the $A(0) f\rho$ as a function of the heliocentric distance along with the fitted slope for the evolution of pre- and post- perihelion activity. From both the gas and the dust production evolution, we see that the comet activity evolves quite regularly. We could not detect any major outburst or sudden change of the comet activity with the heliocentric distance.

\indent
We immediately notice that the five gas species and the dust display an asymmetry about perihelion, since the rate of brightening is steeper than the rate of fading. The asymmetry does not have the same amplitude for the gas and the dust. Indeed, the $A(0) f\rho$ values are higher after perihelion than before at the same heliocentric distance by at least a factor two, while the gas production rates do not display such a strong difference between pre- and post-perihelion values. This of course implies that the dust-to-gas ratio is higher after perihelion.

\begin{table}[h!]
\begin{center}
\caption[Caption for LOF]{Pre- and post-perihelion fitted power-law slopes for $\mathrm{OH}$, $\mathrm{NH}$, $\mathrm{CN}$, $\mathrm{C_{3}}$, and $\mathrm{C_{2}}$ production rates and $A(0) f\rho$ heliocentric dependences.}
\begin{tabular}{lll}
\hline
\hline
  Species  & \multicolumn{2}{c}{$\mathrm{r}$-dependence}\\
           & pre-perihelion     & post-perihelion  \\
\hline
\hline
  OH               & -3.23$\pm$0.33 & -2.62$\pm$0.19 \\
  NH               & -3.55$\pm$0.28 & -2.75$\pm$0.31 \\
  CN               & -2.60$\pm$0.17 & -1.84$\pm$0.09 \\
  $\mathrm{C_{3}}$ & -2.78$\pm$0.40 & -1.75$\pm$0.12 \\
  $\mathrm{C_{2}}$ & -3.56$\pm$0.16 & -2.76$\pm$0.12 \\
  $A(0) f\rho$     & -3.92$\pm$0.19 & -3.01$\pm$0.07 \\                 
\hline
\hline
\end{tabular}
\label{rdependence}
\end{center}
\end{table}

\indent
We adopted a power-law fit to represent the heliocentric dependence of the production rates and the $A(0) f\rho$ both pre- and post-perihelion. The values of the slopes are given in Table \ref{rdependence}. The rate of brightening is steeper than the rate of fading, which confirms the asymmetry about perihelion observed in Figure \ref{gasrh}. The slopes have similar values for OH, NH, and $\mathrm{C_{2}}$ and are consistent within the error bars with the slope observed for the dust heliocentric dependence. However, the slopes for CN and $\mathrm{C_{3}}$ are different from the other gas species and the dust. This difference is visible on both sides of perihelion.

\indent
We have a few observations of comet Lovejoy with the RC and GC continuum filters in addition to the ones regularly performed with the BC filter. The comparison of the  $A(\theta) f\rho$ values in these three filters allows the color of the dust and its evolution
to be studied. The dust color is usually computed as the normalized gradient of $A(\theta) f\rho$ from two continuum filters: 
\begin{equation}
\text{color} [\lambda _{1},\lambda _{2}]=\frac{Af\rho _{1}-Af\rho _{2}}{\lambda _{1}-\lambda _{2}}\frac{2000}{Af\rho _{1}+Af\rho _{2}}
.\end{equation}
The normalized reflectivity gradient is expressed as the percentage of reddening by 1000 \AA. We measured values between 6~$\%$ and 14 $\% /1000$ \AA$ $. This is in the range of what is usually observed for comets (see for example \citealt{Lin2013}; \citealt{Lin2012};  \citealt{Lara2004}). The dust color stayed constant during our observations.

\subsection{Composition}
\label{Compo}

\begin{figure}[h!]
\centering
 \includegraphics[height=6.5cm]{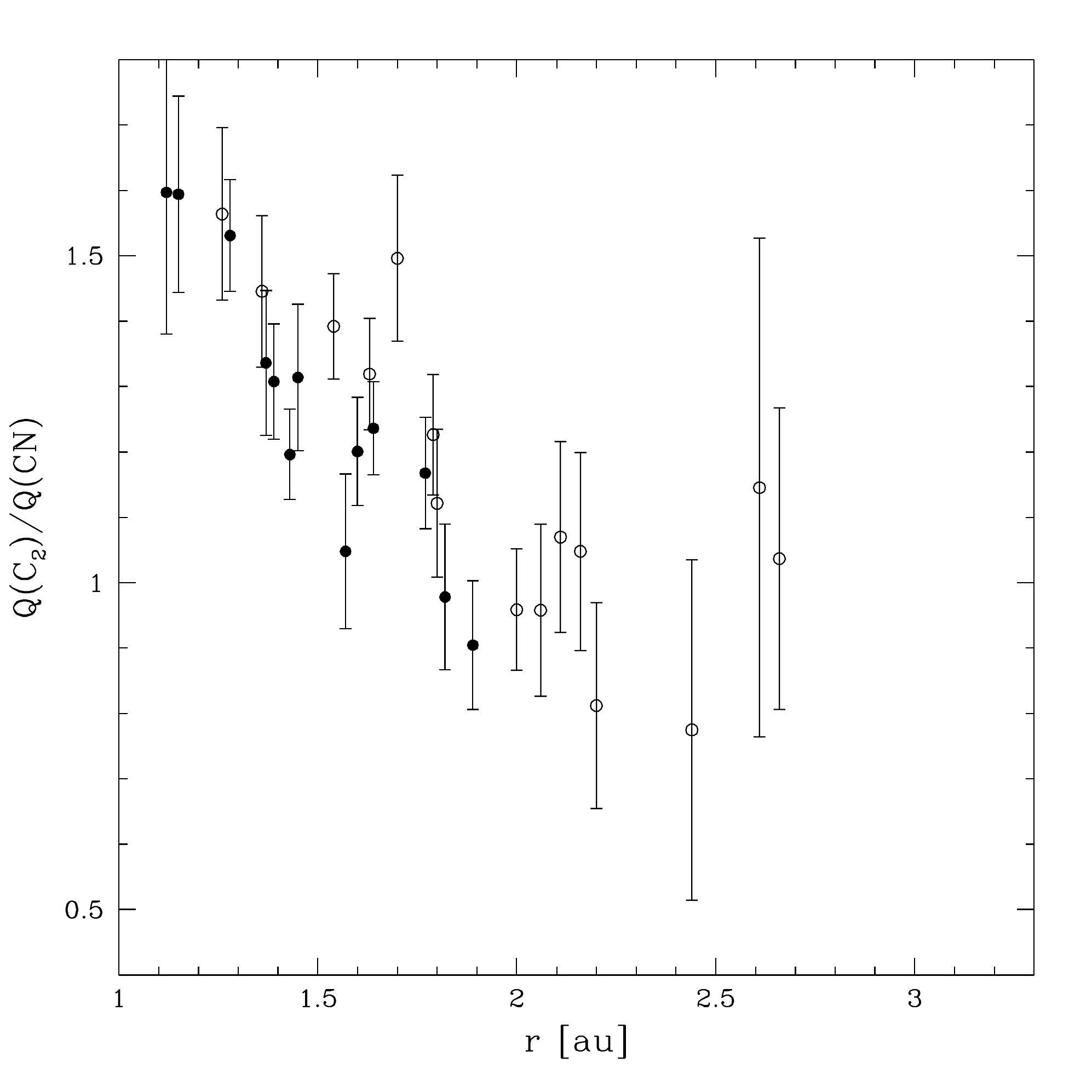}
\caption{Ratio of $\mathrm{C_{2}}$ -to-CN production rates as a function of the heliocentric distance. Pre-perihelion values are represented with filled symbols and post-perihelion values with open symbols.}
  \label{compC}
\end{figure}

\indent
We monitored several production rate ratios to study the evolution of the coma composition. In Fig. \ref{compC} we show the evolution of the $\mathrm{C_{2}}$ -to-CN ratio. This ratio decreases with the heliocentric distance by almost a factor two between 1.1 and 2.5 au. At large heliocentric distances, the ratio does not seem to vary anymore, but the error bars are larger. Pre- and post-perihelion ratios match within the error bars, and both display the same trend. The $\mathrm{C_{2}}$ production rate is rising faster than the CN production rate as the comet approaches the Sun and decreasing faster after perihelion. Despite the evolution of the $\mathrm{C_{2}}$ to CN ratio, comet Lovejoy can be classified as a typical comet in terms of carbon-chain species as defined by \citet{AHearn1995}. However, around 2.5 au, the $\mathrm{C_{2}}$ -to-CN ratio almost reaches the lower limit for a typical comet. This shows that the composition of the coma evolves during the orbit and that a comet could be classified as typical or depleted depending on the heliocentric distance at which it is observed. We must then treat this kind of classification with caution.
\begin{table}[h!]
\begin{center}
\caption[Caption for LOF]{Pre- and post-perihelion fitted power-law slopes for the CN, $\mathrm{C_{3}}$, $\mathrm{C_{2}}$, and $A(0) f\rho$ to OH ratios heliocentric dependences.}
\begin{tabular}{lll}
\hline
\hline
  Species  & \multicolumn{2}{c}{$\mathrm{r}$-dependence}\\
           & pre-perihelion     & post-perihelion  \\
\hline
\hline
  log[Q(CN)/Q(OH)]                 & 1.00$\pm$0.52 & 1.06$\pm$0.20 \\
  log[Q($\mathrm{C_{3}}$)/Q(OH)]   & 1.02$\pm$0.69 & 0.70$\pm$0.30 \\
  log[Q($\mathrm{C_{2}}$)/Q(OH)]   & -0.35$\pm$0.44 & -0.21$\pm$0.23 \\
  log[Q($A(0) f\rho$)/Q(OH)]       & -0.23$\pm$0.36 & -0.48$\pm$0.20 \\                
\hline
\hline
\end{tabular}
\label{slopesrh}
\end{center}
\end{table}

\begin{figure}[h!]
\centering
\subfigure{ \includegraphics[height=8.2cm]{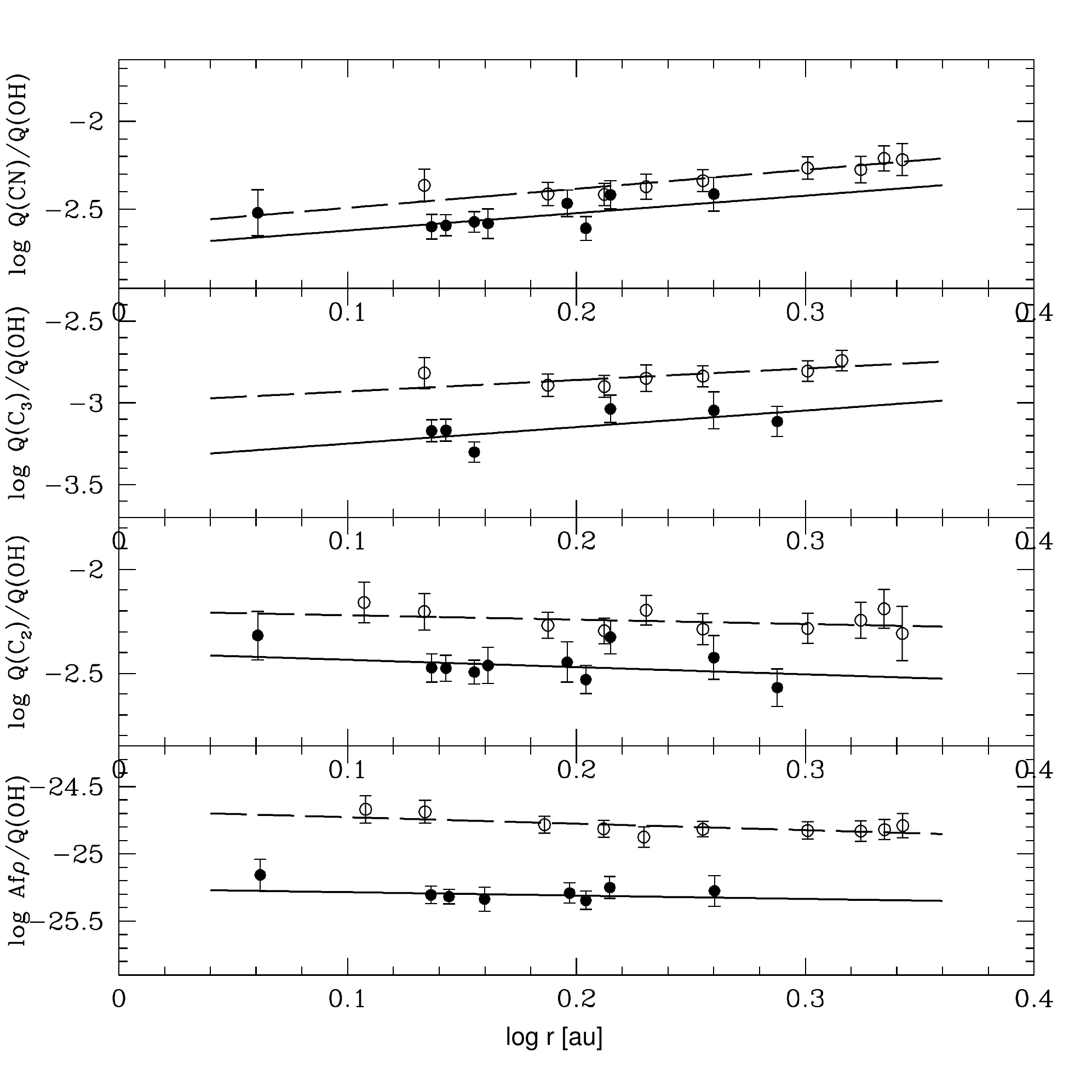}}
\caption{Ratio of CN, $\mathrm{C_{3}}$, $\mathrm{C_{2}}$ production rates, and the $A(0) f\rho$ to the OH production rate as a function of the heliocentric distance. Pre-perihelion values are represented with filled symbols and post-perihelion values with open symbols. Full lines represent linear fits of the ratios variation with the heliocentric distance pre-perihelion. Dotted lines are the same for post-perihelion values.}
  \label{compOH}
\end{figure}

\indent
In Fig. \ref{compOH} we show the evolution of the CN, $\mathrm{C_{3}}$, $\mathrm{and C_{2}}$ production rates, and $A(0) f\rho$ compared to the OH production rate. We did not represent the evolution of the NH/OH ratio since simultaneous measurements of these two species were scarce. The dispersion of the ratios is mainly due to the dispersion of the OH production rates caused by the lower signal-to-noise ratio in these images. We fit two separated slopes for pre- and post-perihelion data, which are given in Table \ref{slopesrh}. All the ratios display a slight asymmetry about perihelion, post-perihelion values are higher than pre-perihelion values owing to the asymmetry of the OH production rates about perihelion. The asymmetry is stronger for $A(0) f\rho$ since, as already noticed, the comet is dustier post-perihelion. Except for the asymmetry about perihelion, the $A(0) f\rho$ -to-OH and $\mathrm{C_{2}}$ -t- OH ratios are almost constant with heliocentric distance. For these two ratios, the slopes given in Table \ref{slopesrh} are approximately consistent with 0. In contrast, the slopes of the CN-to-OH and, to a lesser extent, $\mathrm{C_{3}}$ -to-OH ratios are different from zero. This means that the evolution of $\mathrm{C_{2}}$, OH, and the dust are correlated, while $\mathrm{C_{3}}$ and CN are likely not, as suggested by the analysis of the heliocentric dependences in section \ref{rate}.

\newpage
\subsection{Morphology}
\label{morph}

\indent
In this section, we describe the dust and gas coma morphologies of comet Lovejoy. We tested several techniques to improve the contrast between the coma and the specific features: subtraction and division of an azimuthal median profile, rotational filter, and Larson-Sekanina filter (see \citealt{Schleicher2004}; \citealt{Larson1984}). All these techniques give approximately the same results, but the removal of the azimuthal median profile allows fainter large scale structures to be detected, and we thus present the images obtained with this technique.

\begin{figure*}[htbp!]
\centering
\subfigure{ \includegraphics[width=7.2cm,height=7.2cm]{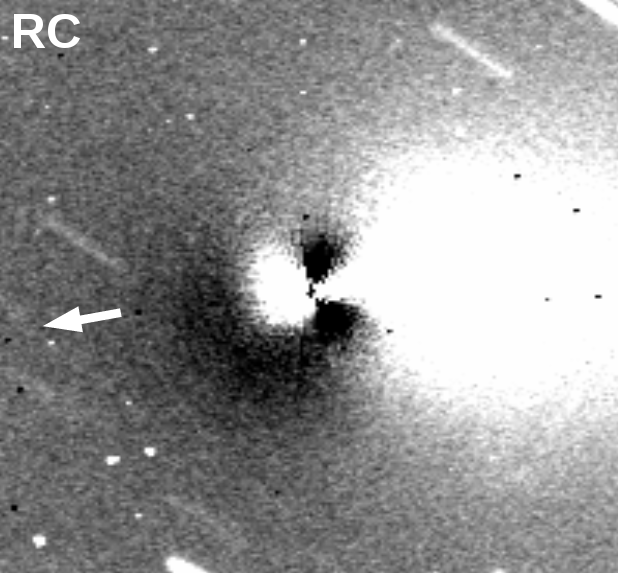}}
\hspace{1cm}
\subfigure{ \includegraphics[width=7.2cm,height=7.2cm]{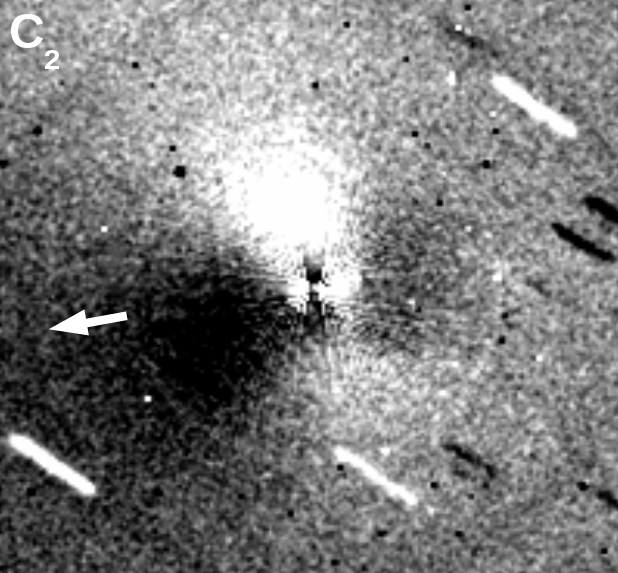}}
\subfigure{ \includegraphics[width=7.2cm,height=7.2cm]{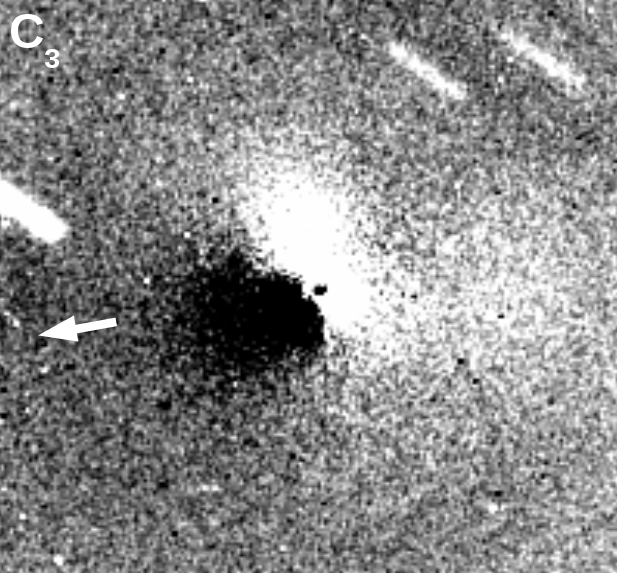}}
\hspace{1cm}
\subfigure{ \includegraphics[width=7.2cm,height=7.2cm]{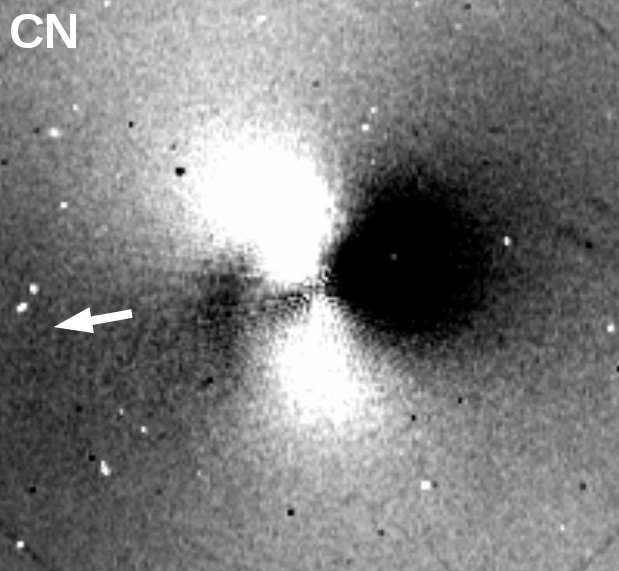}}
\subfigure{ \includegraphics[width=7.2cm,height=7.2cm]{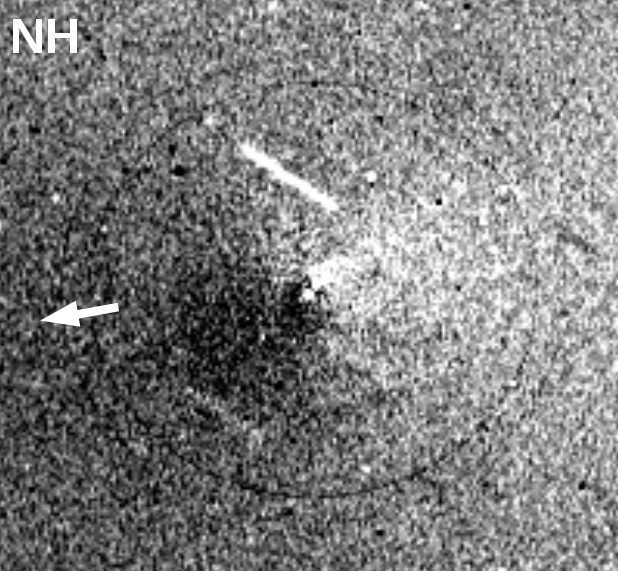}}
\hspace{1cm}
\subfigure{ \includegraphics[width=7.2cm,height=7.2cm]{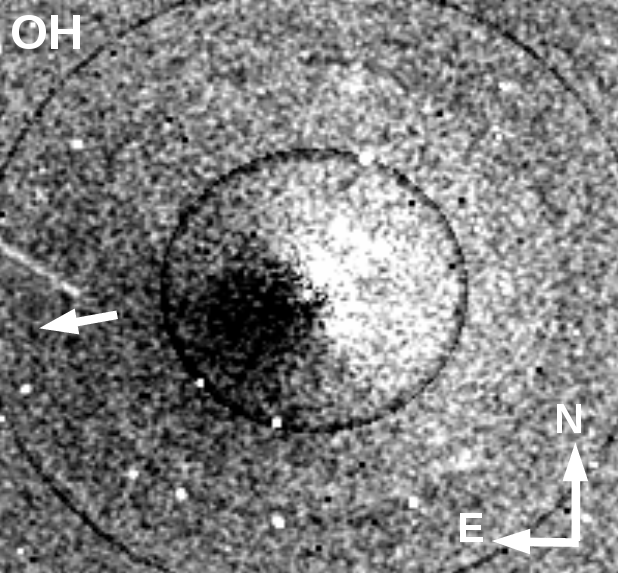}}

\caption{C/2013 R1 (Lovejoy) RC, $\mathrm{C_{2}}$, $\mathrm{C_{3}}$ CN, NH, and OH images from November 2 and 3, 2013 processed by subtracting an azimuthal median profile. All images are oriented with north up and east left. The field of view is 4.3\textquoteright$ $ $\times$ 4.3\textquoteright$ $. The arrow indicates the direction of the Sun.}
  \label{morphology}
\end{figure*}

\indent
We first discuss the observations obtained before perihelion between September 12 and November 16, 2013. We detected features in all the filters, indicating that the nucleus of comet Lovejoy has several active regions. The first CN feature was detected around September 21, 2013, while the dust, $\mathrm{C_{3}}$, and $\mathrm{C_{2}}$, features were only detected later, around October 6, 2013. The OH and NH features do not appear in all the images, owing to the high airmass of some observations and the lower signal-to-noise ratio in these bands. We present in Fig. \ref{morphology} the results of the subtraction of the azimuthal median profile for $\mathrm{C_{2}}$, $\mathrm{C_{3}}$, CN, NH, and OH images, and also for the dust in RC filter for images taken on November 2 and 3, 2013. The RC image shows an enhancement of the coma in the tail direction around PA 280\degre$ $. There is a broad and slightly curved feature centered on PA 100\degre$ $. The $\mathrm{C_{2}}$ image shows two jets around PA 100\degre$ $ and 280\degre$ $, at the same position as the dust features mentioned above. We also see a broad feature around PA 15\degre$ $ and a fainter one around PA 190\degre$ $ that are not visible in the dust images. The features around PA 100\degre$ $ and 280\degre$ $ visible in both $\mathrm{C_{2}}$ and dust images seem to indicate that at least part of the $\mathrm{C_{2}}$ originates in the same region(s) as the dust, or maybe that the $\mathrm{C_{2}}$ is released from dust grains. The CN morphology is different from the $\mathrm{C_{2}}$ morphology. CN displays a roughly hourglass shape with two main fans around PA 15\degre$ $ and 175\degre$ $. Features were also visible in the $\mathrm{C_{2}}$ images at approximately the same position as the CN jets but were much weaker. The $\mathrm{C_{3}}$ morphology is rather similar to the $\mathrm{C_{2}}$ morphology, with one feature around PA 280\degre$ $ and a large fan around PA 5\degre$ $. However, the $\mathrm{C_{2}}$ feature around PA 100\degre$ $ and the weak one around PA 190\degre$ $ are not observed in the $\mathrm{C_{3}}$ image. To summarize, $\mathrm{C_{2}}$ and, to a lesser extent, $\mathrm{C_{3}}$ images are the sum of dust and CN features. In the NH and OH filters, we do not detect jets but only an enhancement of the coma centered on the tail direction, around 280\degre$ $. Since the comet was only visible one or two hours at the end of the night during this period, we could not detect variations in the feature's shape and position during the night caused by the nucleus rotation. The position and the shape of the features described above for November 2 and 3 varied slightly between late September and mid-November because of changes in the viewing geometry. We also have images of the comet post-perihelion. Unfortunately, we could only identify clear features for CN. The morphology is almost the same as in November 2013: an hourglass shape with two main fans around PA 15\degre$ $ and  205\degre$ $.

\section{Discussion}
\label{discussion}
\indent
The study of the evolution of comet Lovejoy activity with the heliocentric distance revealed a significant asymmetry about perihelion. This asymmetry is visible from $\mathrm{C_{2}}$, $\mathrm{C_{3}}$, CN, NH, and OH production rates but also from the dust, even if the amplitude of the asymmetry varies depending on the species. The $A(0) f\rho$ values are at least a factor two higher after perihelion than before at the same heliocentric distance. Asymmetries of the activity about perihelion seem to be a common phenomenon among comets (see, for example, \citealt{AHearn1995}, \citealt{Schleicher1998}, \citealt{Knight2013}, \citealt{Opitom2015}). It remains difficult to know if such asymmetries happen for all comets or to understand their origin, because large datasets with sufficient temporal coverage and with observations on both sides of perihelion are still too scarce. A possible cause often mentioned is a thermal inertia effect. One other probable cause is a seasonal effect on a nucleus with several active regions. Indeed, for comets on highly eccentric orbits, changes in the nucleus illumination can happen close to perihelion. If the illuminated percentage of the active region(s) rises, this will also impact the evolution of the comet gas and dust production rates. Some regions might also need more heat to be activated, for instance if they are buried deeper, or there might be some delay before they start to outgas. The study of comet Lovejoy coma morphology revealed gas and dust jets coming out of the nucleus. This strongly suggests the existence of one or several active region(s) on the nucleus. This is consistent with a seasonal effect being the cause of the observed asymmetry. We also noted that the dust-to-gas ratio changed; the comet is dustier after perihelion than before. This may bring us important clues about the heterogeneity level of the nucleus. More precisely, this could be explained by different ice and dust mixing in the active regions. However, we must be cautious while drawing conclusions since we do not completely understand the mechanism that causes such asymmetries.

\indent
While looking at the evolution of the comet activity with the heliocentric distance, we noticed a correlation between some of the radicals and the dust. The slopes of the $\mathrm{C_{2}}$, NH, OH, and the dust heliocentric dependence presented in Table \ref{rdependence} are consistent with each other within the error bars, while the CN slope is shallower. The $\mathrm{C_{3}}$ slope is close to the CN slope. This correlation between some radicals and the dust is also observed after perihelion. It is even more obvious looking at the evolution of the $\mathrm{C_{2}}$, $\mathrm{C_{3}}$, CN, NH, and dust production rates relative to OH. Except for the asymmetry about perihelion, the $\mathrm{C_{2}}$/OH and $A(0) f\rho$/OH ratios do not significantly vary with heliocentric distance. On the other hand, a power-law fit of the CN/OH  and the $\mathrm{C_{3}}$/OH ratio with the heliocentric distance gives a slope different from zero. Once again, these trends are identical on both sides of perihelion.

\indent
All this evidence suggests that the radicals we observe in the coma of comet Lovejoy may not all come from the same source. The issue of gas coming from icy or carbonaceous grains in the coma of comets has been debated for a long time. $\mathrm{C_{2}H_{2}}$, $\mathrm{C_{2}H_{6}}$, and HCN are usually thought to be the primary parents of $\mathrm{C_{2}}$ and CN (\citealt{Helbert2005}, \citealt{Weiler2012}). However, discrepancies between the parent and daughter production rates observed for some comets seem to indicate that we need other parents to explain the abundances of some radicals in the coma (\citealt{Fray2005}; \citealt{DelloRusso2009}; \citealt{Kobayashi2010};  \citealt{McKay2014}). It has been suggested that $\mathrm{C_{2}}$ and/or CN can come -at least partially- from carbonaceous grains or another unknown source. In the case of comet Lovejoy, we see a strong correlation between the evolution of the dust production and the evolution of some radicals such as $\mathrm{C_{2}}$, NH, and OH. This suggests that at least part of the $\mathrm{C_{2}}$, on one hand, and also part of the NH and OH, on the other hand, could be produced by organic-rich and icy grains, respectively. Since the evolution of the CN production is different from the $\mathrm{C_{2}}$ and does not seem correlated with the dust, most CN would be produced directly by the photodissociation of HCN. 

\indent
This hypothesis is further reinforced by the study of the coma morphology. Indeed, Fig. \ref{morphology} shows that the dust and $\mathrm{C_{2}}$ morphologies have some similarities. $\mathrm{C_{2}}$ is enhanced in the tail direction, and we see a jet in the sunward direction at the same position as the dust jet. We also notice features around the same position as the CN fans. In contrast, the CN morphology is completely different from the dust. It has an hourglass shape with two jets around PA 15\degre$ $ and 175\degre$ $. The $\mathrm{C_{3}}$ morphology is similar to the $\mathrm{C_{2}}$ morphology. Since HCN, $\mathrm{C_{2}H_{2}}$, and $\mathrm{C_{2}H_{6}}$ usually have the same spatial distribution (\citealt{DelloRusso2009}, \citealt{Villanueva2011}), this could indicate that, while most of the CN is the product of HCN photodissociation, only part of the $\mathrm{C_{2}}$ is produced by the photodissociation of $\mathrm{C_{2}H_{2}}$ (and $\mathrm{C_{2}H_{6}}$ but $\mathrm{C_{2}H_{6}}$ is not very efficient to release $\mathrm{C_{2}}$, see \citealt{Weiler2012}). The rest of the $\mathrm{C_{2}}$ is probably coming from organic-rich grains. This would explain the enhanced $\mathrm{C_{2}}$ regions around PA 15\degre$ $ and 175,\degre$ $ while the most contrasted features are similar to the dust. We mentioned earlier that NH and OH parents could be released from icy grains. If this is the case, we would expect the NH and OH coma to be enhanced in the tail direction since the icy grains are pushed away from the Sun by the radiation pressure. This is exactly what we see in Fig. \ref{morphology}. Both NH and OH are enhanced in the anti-solar direction, strongly reinforcing the hypothesis of water and ammonia (the most likely parent of NH) being released by icy grains.

\begin{table*}
\centering
\begin{threeparttable}[htbp!]
        \caption{Comparison of mother and daughter species production rates.}
        \label{compa}
        \centering
\begin{tabular}{lccccccr}
\hline 
Date  & Q(OH)\tnote{a}      & Q($\mathrm{H_{2}O}$)\tnote{b} & Q(CN)\tnote{a} & Q(HCN)\tnote{b} & Q($\mathrm{C_{2}}$)\tnote{a} & Q($\mathrm{C_{2}H_{2}}$)\tnote{b} & Q($\mathrm{C_{2}H_{6}}$)\tnote{b}  \\ 
                  
\hline
2013-Oct-21    & 2380$\pm$250 &                & 6.38$\pm$0.19  &                       & 7.63$\pm$0.21  &            &                  \\  
2013-Oct-22    &              &  1850$\pm$190  &                &  5.9$\pm$1.2          &                &   $<$9.2   &  13.80$\pm$1.70  \\
2013-Oct-24    & 2720$\pm$290 &  1530$\pm$250  & 6.96$\pm$0.21  &                       & 9.10$\pm$0.34  &            &                  \\
2013-Oct-25    &              &  2320$\pm$330  &                &  6.8$\pm$0.6          &                &  $<$1.5    &  21.70$\pm$3.50  \\
2013-Oct-26    & 2860$\pm$330 &                & 7.20$\pm$0.31  &                       & 9.62$\pm$0.39  &            &                  \\
2013-Oct-27    &              &  1560$\pm$250  &                &  3.6$\pm$0.1          &                &  $<$2.2    &  17.40$\pm$3.70  \\
2013-Oct-29    &              &  1450$\pm$410  &                &  4.7$\pm$0.6          &                &  $<$1.7    &  10.80$\pm$2.50   \\
2013-Nov-2     &              &                & 7.97$\pm$0.27  &                       & 12.20$\pm$0.30 &            &                   \\
2013-Nov-3     & 3010$\pm$350 &                &                &                       &                &            &                   \\  
2013-Nov-7     &              &  3680$\pm$260  &                &  8.7$\pm$0.7          &                &  $<$1.6    &  21.80$\pm$2.50   \\  
2013-Nov-8/12  &              & \tnote{d} 5000 &                & \tnote{d} 7.4$\pm$0.1 &                &            &                   \\ 
2013-Nov-13    & 3024$\pm$790 &                & 10.10$\pm$0.70 & \tnote{c} 6.0         & 16.10$\pm$0.50 &            &                   \\  
2013-Nov-16    &              &                & 12.90$\pm$1.20 & \tnote{c} 6.1         & 20.60$\pm$0.90 &            &                   \\ 
\hline 
\end{tabular}
        \begin{tablenotes}
                \footnotesize
                  \item All production rates are expressed in units of $10^{25}$ mol/s.
                  \item[a] This paper
                  \item[b] \cite{Paganini2014b}, from observations with the Near InfraRed SPECtrometer at the Keck observatory
                \item[c] \cite{Agundez2014}, from Nancay OH observations
                \item[d] \cite{Biver2014}, from observations with the IRAM 30m telescope
        \end{tablenotes}
\end{threeparttable}
\end{table*}

\indent
To better understand the relationship between parent and daughter species,  we can compare measurements of the parent species production rates made in the IR and the radio domains to the radicals production rates we derived here. Only a few results have been published so far about comet C/2013 R1 (Lovejoy). We summarize the results reported in \cite{Paganini2014b}, \cite{Agundez2014}, and \cite{Biver2014} in Table \ref{compa}, along with our daughter species production rates. From this table, we notice that there is a significant dispersion between production rates measured almost simultaneously with different instruments and from different wavelength ranges. This dispersion is probably due in part to the different fields of view (less than 10 arcseconds in the millimeter range and 24 arcseconds in the IR) and models used but also maybe to the rotational variation.

\indent
We first compare the OH and $\mathrm{H_{2}O}$ production rates. Our values agree within the error bars with the ones reported by \cite{Biver2014} and \cite{Paganini2014b} for early November. In October, however, the OH production rates we derived are slightly higher than the water production rates measured by \cite{Paganini2014}. 

\indent
Since measurements of the HCN production rate of comet Lovejoy have also been published by \cite{Paganini2014b} and \cite{Agundez2014}, the comparison with our CN measurements is of great interest for determining the origin of the CN in the coma of comet Lovejoy. In October, the CN and HCN production rates are consistent between each other (except for October 27 and 29 for which HCN production rates reported by \citet{Paganini2014b} are almost a factor two lower than those measured a few days before and a few days after, maybe because of rotational variation very close to the nucleus). In November, our CN production rates agree with those reported by \cite{Paganini2014b} and \cite{Biver2014}, especially if we take the rise in comet activity into account. The HCN production rates reported by \cite{Agundez2014} are lower than our CN production rates, but they are also lower than those derived by \cite{Paganini2014b} and \cite{Biver2014} only a few days earlier.
 
\indent
Finally, we can also compare the $\mathrm{C_{2}}$ production rates to the production rates of two possible parents, $\mathrm{C_{2}H_{6}}$ and $\mathrm{C_{2}H_{2}}$, reported by \cite{Paganini2014b}. For all dates except one, the $\mathrm{C_{2}H_{2}}$ production rate upper limit is lower than the $\mathrm{C_{2}}$ production rate. For some dates the $\mathrm{C_{2}H_{2}}$ production rate upper limit is almost a factor ten lower than the $\mathrm{C_{2}}$ production rate. The $\mathrm{C_{2}H_{6}}$ is of the same order of magnitude as the $\mathrm{C_{2}}$ production rate.

\indent
The comparison of parent and daughter species production rates seems to confirm our interpretation of the origin of daughter species in the coma of comet Lovejoy. Most of the HCN production rates reported are of the same order of magnitude as our CN production rates, the ones reported by \cite{Paganini2014b} are even in good agreement with our CN production rates. This suggests that HCN is probably the main parent of CN in the coma of comet Lovejoy. For $\mathrm{C_{2}}$, there are large discrepancies between the most likely parent abundance and the $\mathrm{C_{2}}$ abundance. Indeed most of the upper limits of the $\mathrm{C_{2}H_{2}}$ production rates are much lower than the $\mathrm{C_{2}}$ production rates. This indicates that we need another parent to explain the $\mathrm{C_{2}}$ abundance in the coma or that $\mathrm{C_{2}}$ comes in part from organic-rich grains. Another possible parent of $\mathrm{C_{2}}$ is $\mathrm{C_{2}H_{6}}$. \cite{Paganini2014b} detected $\mathrm{C_{2}H_{6}}$ in the coma of comet Lovejoy, and its abundance is of the same order of magnitude as the $\mathrm{C_{2}}$. However, it has been reported that the release of $\mathrm{C_{2}}$ from $\mathrm{C_{2}H_{6}}$ is not really efficient (\citealt{Weiler2012}) and makes it a poor candidate for $\mathrm{C_{2}}$ parentage.

\indent
Nevertheless, we must remain cautious when comparing parent molecules production rates measured from the infrared to daughter molecules production rates from the optical. The models used and the parameters in these models are different, making any direct comparison difficult. As an example, in this paper we used an outflow velocity of 1 km/s, as in \cite{AHearn1995}. However, in most infrared and radio studies, including those from \cite{Paganini2014b} and \cite{Biver2014} mentioned earlier, an outflow velocity of $\mathrm{v}=0.8r^{-0.5}$ km/s is assumed. To assess the importance of the outflow velocity used in the model on the derived daughter molecules production rates and then on the comparison between daughter and parent species production, we computed the gas production rates presented in Table \ref{compa} again using the expansion velocity law: $\mathrm{v}=0.8r^{-0.5}$ km/s. These results are shown in Table \ref{compabis}. A comparison of Tables \ref{compa} and \ref{compabis} shows that the impact of the outflow velocity is significant. The production rates derived from optical data are lower if we use the same velocity law as for the infrared data (lower by a factor 1.8 in October and by a factor 1.5 in mid-November). However, the agreement between the observed profile and the Haser model is much better for constant expansion velocities of 1 km/s. This is not surprising since the Haser scalelengths we use \citep{AHearn1995} have been fitted using an expansion velocity of 1 km/s. 

\indent

Examination of Table \ref{compabis} shows that given the variability of water production rates measurements, there are no significant differences between OH production rates and those of $\mathrm{H_{2}O}$ for most of the dates. Some of the CN production rates measured are lower than those of HCN reported in the literature, but they are consistent with the lower HCN production rates measured in late October and mid-November. In early November, $\mathrm{C_{2}}$ production rates are still almost a factor five higher than the $\mathrm{C_{2}H_{2}}$ production rates upper limits.

\small
\begin{table*}
\centering
\begin{threeparttable}[hbp!]
        \caption{Comparison of mother and daughter species production rates using a gas outflow velocity $v=0.8r^{-0.5}$ km/s to compute daughter species production rates.}
        \label{compabis}
        \centering
\begin{tabular}{lccccccr}
\hline 
Date  & Q(OH)\tnote{a}      & Q($\mathrm{H_{2}O}$)\tnote{b} & Q(CN)\tnote{a} & Q(HCN)\tnote{b} & Q($\mathrm{C_{2}}$)\tnote{a} & Q($\mathrm{C_{2}H_{2}}$)\tnote{b} & Q($\mathrm{C_{2}H_{6}}$)\tnote{b}  \\ 
                  
\hline
2013-Oct-21    & 1280$\pm$130 &                & 3.53$\pm$0.11  &                       & 4.20$\pm$0.12  &            &                  \\  
2013-Oct-22    &              &  1850$\pm$190  &                &  5.9$\pm$1.2          &                &   $<$9.2   &  13.80$\pm$1.70  \\
2013-Oct-24    & 1500$\pm$160 &  1530$\pm$250  & 3.95$\pm$0.12  &                       & 5.15$\pm$0.19  &            &                  \\
2013-Oct-25    &              &  2320$\pm$330  &                &  6.8$\pm$0.6          &                &  $<$1.5    &  21.70$\pm$3.50  \\
2013-Oct-26    & 1610$\pm$190 &                & 4.16$\pm$0.18  &                       & 5.52$\pm$0.22  &            &                  \\
2013-Oct-27    &              &  1560$\pm$250  &                &  3.6$\pm$0.1          &                &  $<$2.2    &  17.40$\pm$3.70  \\
2013-Oct-29    &              &  1450$\pm$410  &                &  4.7$\pm$0.6          &                &  $<$1.7    &  10.80$\pm$2.50   \\
2013-Nov-2     &              &                & 4.90$\pm$0.17  &                       & 7.46$\pm$0.16 &            &                   \\
2013-Nov-3     & 1810$\pm$210 &                &                &                       &                &            &                   \\  
2013-Nov-7     &              &  3680$\pm$260  &                &  8.7$\pm$0.7          &                &  $<$1.6    &  21.80$\pm$2.50   \\  
2013-Nov-8/12  &              & \tnote{d} 5000 &                & \tnote{d} 7.4$\pm$0.1 &                &            &                   \\ 
2013-Nov-13    & 2090$\pm$510 &                &  6.88$\pm$0.44 & \tnote{c} 6.0         & 10.90$\pm$0.40 &            &                   \\  
2013-Nov-16    &              &                &  8.97$\pm$0.84 & \tnote{c} 6.1         & 14.30$\pm$0.60 &            &                   \\ 
\hline 
\end{tabular}
        \begin{tablenotes}
                \footnotesize
                  \item All production rates are expressed in units of $10^{25}$ mol/s.
                  \item[a] This paper
                  \item[b] \citet{Paganini2014b}, from observations with the Near InfraRed SPECtrometer at the Keck observatory
                \item[c] \citet{Agundez2014}, from Nancay OH observations
                \item[d] \citet{Biver2014}, from observations with the IRAM 30m telescope
        \end{tablenotes}
\end{threeparttable}
\end{table*}
\normalsize

\indent
Another question is the dependence of both scalelengths and fluorescence efficiencies on the heliocentric distance. This of course influences the daughter production rates we derive and then the comparison between daughter and parent molecules production. In our case, we only compare daughter and parent molecules for observations made between 1.4 and 1 au from the Sun so that variations from the $r^{2}$ scaling used for both scalelengths and g~factors should not have a strong influence on our results. For example, if we use a $r^{3}$ scaling law for the scalelengths, the maximum variation in the production rates over this range heliocentric distances is 20$\%$. Finally, the $\mathrm{C_{2}}$ fluorescence efficiencies we used only consider the $\mathrm{C_{2}}$ in triplet state since the ratio of $\mathrm{C_{2}}$ in triplet and singlet state and its heliocentric dependence are not understood well. We thus underestimate the $\mathrm{C_{2}}$ total production rates, which reinforces our conclusion about $\mathrm{C_{2}}$ parentage.

\section{Summary}
\label{summary}
\indent
We performed a photometry and imaging monitoring of the Oort Cloud comet C/2013 R1 (Lovejoy) during several months with the robotic TRAPPIST telescope at La Silla observatory. We observed the comet on both sides of perihelion with good temporal coverage. We derived production rates and $Af\rho$ values and studied the evolution of its activity. We also performed a morphological analysis of the coma. 

\indent
The morphological analysis revealed several features in the coma, both for the dust and the gas species. This indicates that the nucleus has one or several active region(s). The comparison of the comet activity pre- and -post perihelion shows an asymmetry of the gas and dust production rates about perihelion, the rate of brightening is steeper than the rate of fading. We estimated that it is likely to be seasonal effect caused by changing system geometry and insolation of the active region(s). The dust production rate is at least two times higher after perihelion than before at the same heliocentric distance, which suggests a certain degree of heterogeneity of the nucleus.

\indent
The evolution of the comet activity and production rates ratios with the heliocentric distance, the comparison of parent and daughter species production rates, as well as the coma morphology all indicate that the $\mathrm{C_{2}}$, NH, and OH productions are strongly correlated to the dust production, while the CN production is not. The case of $\mathrm{C_{3}}$ is more complicated and quite intermediate between these two behaviors. We then suggested that an important part of the $\mathrm{C_{2}}$ could be released from organic-rich grains and that water and ammonia (the parents of OH and NH) could be released from icy grains. CN and some $\mathrm{C_{2}}$ would be produced by the photodissociation of HCN and $\mathrm{C_{2}H_{2}}$, which usually display the same spatial distribution. $\mathrm{C_{3}}$ would also be produced partly from the organic-rich grains and by the photodissociation of a parent molecule.

\begin{acknowledgements}
TRAPPIST is a project funded by the Belgian Fund for Scientific Research (Fonds National de la Recherche Scientifique, F.R.S.-FNRS) under grant FRFC 2.5.594.09.F, with the participation of the Swiss National Science Fundation (SNF). C. Opitom acknowledges the support of the FNRS. E. Jehin and M. Gillon are FNRS Research Associates, D. Hutsem\'ekers is FNRS Senior Research Associate. We are grateful to David Schleicher and the Lowell Observatory for the loan of a set of NASA HB comet filters. We thank Claude Arpigny and Philippe Rousselot for helpful discussions.
\end{acknowledgements}


\bibliographystyle{aa}
\bibliography{Biblio}

\end{document}